\documentclass[a4paper]{article}
\usepackage{epsfig}
\newcommand{\e}{\mathrm{e}}
\newcommand{\ep}{\epsilon}
\newcommand{\vev}[1]{\left\langle #1 \right\rangle}
\newcommand{\V}{\mathcal{V}}
\begin{document}
\title{\textbf{Integral Equations for ERG}}
\author{Hidenori Sonoda\footnote{\texttt{hsonoda@kobe-u.ac.jp}}\\
\small \textit{Physics Department, Kobe University, Kobe 657-8501, Japan}}
\date{November 2005}
\maketitle

\begin{abstract}
An application of the exact renormalization group equations to the
scalar field theory in three dimensional euclidean space is
discussed.  We show how to modify the original formulation by
J.~Polchinski in order to find the Wilson-Fisher fixed point using
perturbation theory.
\end{abstract}

\section{Introduction}

The exact renormalization group (ERG) equation, first introduced by
K.~Wilson \cite{Wilson:1973jj}, was reformulated by J.~Polchinski
\cite{Polchinski:1983gv} in a form more suitable for perturbation
theory.  In this paper we apply Polchinski's ERG differential
equations to three dimensional scalar field theory.

In the first half of the paper (sects.~2 and 3), we examine the nature
of the solutions to the ERG differential equations.  We introduce
integral equations that combine the differential equations and the
asymptotic conditions.  The latter specify the solution unambiguously,
and the integral equations, under a given set of parameters, have a
unique solution.  The main issue is self-similarity of the solutions:
whether a change of the renormalization scale can be compensated by
changing the parameters of the solutions.  We show it is impossible to
have self-similarity unless we keep an unphysical parameter.  In
sect.~4, we will modify the ERG differential equations to acquire
self-similarity without any unphysical parameter.  Finally, in
sect.~5, we modify the ERG differential equations further so that the
Wilson-Fisher fixed point is accessible by perturbation theory.

Note that in order to have fixed points of ERG, it is necessary to
rescale the momenta so that the renormalization momentum scale, say
$\mu$, remains fixed.  Throughout the paper we adopt the convention
\begin{equation}
\mu = 1
\end{equation}

\section{First rewriting: integral equations \label{integral}}

The action is given as
\begin{eqnarray}
S (t) &=& \frac{1}{2} \int_p \phi (p) \phi (-p) \frac{p^2 + m^2
\e^{2t}}{K(p)} \nonumber\\ && - \sum_{n=1}^\infty \frac{1}{(2n)!}
\int_{p_1 + \cdots + p_{2n} = 0} \phi (p_1) \cdots \phi (p_{2n})\,
\V_{2n} (t; p_1, \cdots, p_{2n})
\end{eqnarray}
where we use the notation
\begin{equation}
\int_p \equiv \int \frac{d^3 p}{(2 \pi)^3},\quad
\int_{p_1 + \cdots + p_{2n}=0} \equiv \int \prod_{i=1}^{2n} \frac{d^3
  p_i}{(2 \pi)^3} \, (2 \pi)^3 \delta^{(3)} (\sum_{i=1}^{2n} p_i)
\end{equation}
The cutoff function $K(q)$ is a decreasing positive function of $q^2$
with the properties
\begin{equation}
K(q) = \left\lbrace\begin{array}{c@{\quad}l}
 1 & (q^2 < 1)\\
 0 & (q^2 \to \infty)\end{array}\right.
\end{equation}
We also define 
\begin{equation}
\Delta (q) \equiv - 2 q^2 \frac{d}{dq^2} K(q)
\end{equation}
which vanishes for $q^2 < 1$ and is positive for $q^2 > 1$.

The correlation functions of the scalar field are calculated
perturbatively in terms of the propagator
\begin{equation}
\frac{K(p)}{p^2 + m^2 \e^{2t}}
\end{equation}
and the vertices $\{\V_{2n} (t; p_1, \cdots, p_{2n})\}$.  We must
introduce specific $t$-dependence to the vertices $\{\V_{2n} (t)\}$ so
that\footnote{To be precise, this relation is valid only if $p_i^2 <
\e^{- 2t}$ for all $i$.}
\begin{equation}
\vev{\phi (p_1 \e^t) \cdots \phi (p_{2n} \e^t)}_{m^2 \e^{2t}; \V_{2n}
  (t)} = \e^{(y_{2n} - 4 n) t} \vev{\phi (p_1) \cdots \phi
  (p_{2n})}_{m^2; \V_{2n} (0)}
\end{equation}
where 
\begin{equation}
y_{2n} \equiv 3 - n
\end{equation}
is the canonical scale dimension of the $2n$-point vertex $\V_{2n}$.
We note that due to rescaling of momenta under renormalization, the
momenta grow as $\e^t$, and the squared mass grows as $\e^{2t}$.

The above equality is satisfied if the vertices satisfy the following
ERG differential equations, first derived in
\cite{Polchinski:1983gv}\footnote{$\e^{- y_{2n} t} \V_{2n} (t; p_1
  \e^t, \cdots)$ should be replaced by $\V_{2n} (t; p_1, \cdots)$, if
  we do not rescale the momenta under renormalization.}:
\begin{eqnarray}
&&\frac{\partial}{\partial t} \left( \e^{- y_{2n} t} \V_{2n} (t; p_1
\e^t, \cdots, p_{2n} \e^t )\right) \nonumber\\
&=& \sum_{k=0}^{\left[\frac{n-1}{2}\right]} \sum_{i}
\e^{- y_{2(k+1)} t} \V_{2(k+1)} (t; p_{i_1} \e^t, \cdots, p_{i_{2k+1}}
\e^t, (\underbrace{p_{i_{2(k+1)}} + \cdots + p_{i_{2n}}}_{\equiv p})
\e^t)\nonumber \\
&& \qquad 
\times \frac{\Delta (p \e^t)}{p^2 + m^2} \cdot \e^{- y_{2(n-k)} t} \V_{2(n-k)}
(t;  - p \e^t, p_{i_{2(k+1)}} \e^t, \cdots, p_{i_{2n}} \e^t)\nonumber\\
&& + \frac{1}{2} \int_q \frac{\Delta (q \e^t)}{q^2 + m^2} \,
\e^{- y_{2(n+1)} t} \V_{2(n+1)} 
(t; q \e^t, -q \e^t, p_1 \e^t, \cdots, p_{2n} \e^t)
\end{eqnarray}
where the index $i$ runs over the partitions of $2n$ momenta into two
groups.  

We will often encounter the right-hand side of the above equation in
the rest of this paper.  It will save a lot of writing if we introduce
a graphical notation.  By denoting a vertex $\e^{- y_{2n} t} \V_{2n}
(t)$ by
\begin{center}
\epsfig{file=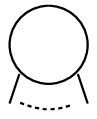}
\end{center}
and $\frac{\Delta (p \e^t)}{p^2 + m^2}$ by a thick line
\begin{center}
\epsfig{file=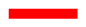}
\end{center}
we can rewrite the ERG differential equations as
\begin{center}
\parbox{0.5cm}{$\displaystyle \frac{\partial}{\partial t}$}
\parbox{1.5cm}{\epsfig{file=vertex.eps}}
\parbox{2cm}{$\displaystyle = \sum_{\mathrm{partitions}}$}
\parbox{2.5cm}{\epsfig{file=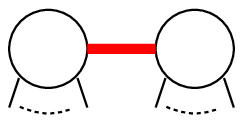}}
\parbox{1cm}{$\displaystyle + \frac{1}{2} \int_q$}
\parbox{2cm}{\epsfig{file=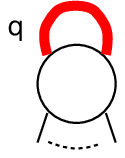}}
\end{center}

The solutions of the ERG differential equations that originate from
the trivial UV fixed point at $t = - \infty$ can be characterized
completely by their asymptotic behaviors as $t \to - \infty$:
\begin{eqnarray}
\e^{- 2 t} \V_2 (t; p \e^t, - p \e^t) &\stackrel{t \to -
  \infty}{\longrightarrow}& \lambda \e^{-t} a_2 + \lambda^2 ( - C t +
  b_2 ) \\ \e^{- t} \V_4 (t; p_1 \e^t, \cdots, p_4 \e^t) &\stackrel{t
  \to - \infty}{\longrightarrow}& - \lambda\\ \e^{- y_{2n} t} \V_{2n
  \ge 6} (t; p_1 \e^t, \cdots, p_{2n} \e^t) &\stackrel{t \to -
  \infty}{\longrightarrow}& 0
\end{eqnarray}
where the two constants $\lambda$, $b_2$ are arbitrary, but $a_2$ and
$C$ are determined uniquely.  The constant $\lambda > 0$ is of course
the coupling constant, while $b_2$ only shifts the origin of the
logarithmic parameter $t$ and is expected to be unphysical.  Hence,
the vertices $\V_{2n} (t; p_1, \cdots, p_{2n})$ depend on three
arbitrary parameters\footnote{We fix the choice of the cutoff function
$K$.}:
\begin{enumerate}
\item $m^2$ which appears in the ERG differential equations
\item $\lambda > 0$ which determines the asymptotic behavior of $\V_4$
\item $b_2$ which determines the asymptotic behavior of $\V_2$
\end{enumerate}
When we wish to show the parametric dependence explicitly, we will use
the following notation
\[
\V_{2n} (t; p_1, \cdots, p_{2n}; m^2, \lambda, b_2)
\]

Before introducing integral equations, let us make a simple
observation.  If the vertices $\{\V_{2n} (t; p_1, \cdots, p_{2n})\}$
solve the ERG differential equations for the squared mass $m^2$, it is
straightforward to show that the shifted vertices $\{\V_{2n} (t +
\Delta t; p_1, \cdots, p_{2n})\}$ also satisfy the ERG differential
equations for the squared mass $m^2 \e^{2 \Delta t}$.  Examining the
asymptotic behaviors of the shifted vertices, we conclude
\begin{eqnarray}
&&\V_{2n} (t + \Delta t; p_1, \cdots, p_{2n}; m^2, \lambda,
  b_2)\nonumber\\
&=& \V_{2n} (t; p_1, \cdots, p_{2n}; m^2 e^{2 \Delta t}, \lambda
\e^{\Delta t}, b_2 - C \Delta t) \label{shift}
\end{eqnarray}
We call this property \textbf{self-similarity} following the standard
nomenclature, meaning that a shift of the logarithmic scale variable
$t$ can be absorbed by changes of the parameters of the vertices.
(Fig.~1.)  The above shows that it is essential to keep $b_2$ if we
wish to have self-similarity.  However, as we will show more
explicitly in the next section, $b_2$ is an unphysical parameter.  If
we wish to have self-similarity with only two parameters $m^2$ and
$\lambda$, we need to modify the ERG differential equations
themselves. This will be discussed in sect.\ref{self-similarity}.
\begin{figure}
\begin{center}
\epsfig{file=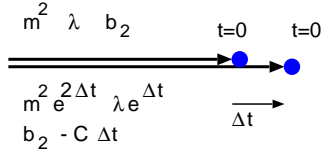}
\caption{The ERG trajectory specified by $m^2 \e^{2 \Delta t}, \lambda
  \e^{\Delta t},~b_2 - C \Delta t$ is the same as the one specified by
  $m^2, \lambda,~b_2$.  But the parameter $t$ is shifted by $\Delta t$.}
\end{center}
\end{figure}

Now, to compute the vertices perturbatively, it is convenient to
convert the ERG \textbf{differential} equations into \textbf{integral}
equations that incorporate the asymptotic behaviors explicitly.  The
integral equations of this type have been discussed extensively for
the four dimensional scalar theory in ref.~\cite{Sonoda:2002pb}, and
we merely transpose the results to three dimensions.  For the
two-point vertex we obtain
\begin{eqnarray}
&& \e^{- 2 t} \V_2 (t; p \e^t, - p \e^t) \nonumber\\
&=& \int_{-\infty}^t dt' \Bigg[\, \e^{- 2 t'} \V_2 (t'; p \e^{t'}, - p
    \e^{t'}) \frac{\Delta (p \e^{t'})}{p^2 + m^2} \e^{- 2 t'} \V_2
    (t'; p \e^{t'}, - p \e^{t'})\nonumber\\
&& + \frac{1}{2} \int_q \frac{\Delta (q \e^{t'})}{q^2 + m^2}
    \, \e^{- t'} \V_4 (t'; q \e^{t'}, - q \e^{t'}, p \e^t, - p
    \e^t) + \e^{-t'} \lambda a_2 + \lambda^2 C \, \Bigg]\nonumber\\
&& + \e^{-t} \lambda a_2 + \lambda^2 ( - C t + b_2 )
\end{eqnarray}
The integral over $t'$ is convergent thanks to the UV subtraction.  To
compensate for the unwanted $t$ dependence of the subtraction, we must
introduce finite counterterms.  The parameter $b_2$ enters as an
integration constant.  For the four-point vertex, we obtain
\begin{eqnarray}
&&\e^{-t} \V_4 (t; p_1 \e^t, \cdots, p_4 \e^t) \nonumber\\
&=& \int_{-\infty}^t dt' \Bigg[ \, \sum_{i=1}^4 \e^{- 2t'} \V_2 (t';
    p_i \e^{t'}, - p_i \e^{t'}) \frac{\Delta (p_i \e^{t'})}{p_i^2 +
      m^2} \cdot \e^{- t'} \V_4 (t'; p_1 \e^{t'}, \cdots, p_4
    \e^{t'})\nonumber\\
&& + \frac{1}{2} \int_q \frac{\Delta (q \e^{t'})}{q^2 + m^2} \,
\V_6 (t'; q \e^{t'}, - q \e^{t'}, p_1 \e^{t'}, \cdots, p_4 \e^{t'}) \,
\Bigg] - \lambda
\end{eqnarray}
The integral over $t'$ is convergent.  The coupling $\lambda$ is
introduced as an integration constant.  For the six-point and higher
vertices, we obtain

\vspace{0.5cm}
\parbox{5cm}{\hfill$\displaystyle 
\e^{- y_{2n} t} \V_{2n \ge 6} (t; p_1 \e^t, \cdots, p_{2n} \e^t)$}\\
\parbox{3.5cm}{$\displaystyle = \int_{-\infty}^t dt' \Bigg[
\, \sum_{\mathrm{partitions}}$}
\parbox{3cm}{\epsfig{file=erg1.eps}}
\begin{equation}
 + \frac{1}{2} \int_q \frac{\Delta (q \e^{t'})}{q^2 + m^2}\,
 \e^{- y_{2(n+1)} t'} \V_{2(n+1)} (t'; q \e^{t'}, - q \e^{t'}, p_1
 \e^{t'}, \cdots, p_{2n} \e^{t'}) \: \Bigg]
\end{equation}
No integration constant is necessary.  In the above integral
equations, the convergence of the $t'$ integral guarantees the
expected asymptotic behaviors.

Solving the integral equations recursively in powers of $\lambda$, we
obtain
\begin{eqnarray}
a_2 &=& \frac{1}{2} \int_q \frac{\Delta (q)}{q^2}\label{a2}\\
C &=& - \frac{1}{2} \int_q \frac{\Delta (q)}{q^2} \int_r \frac{1 -
  K(r)}{r^2} \frac{1 - K(q+r)}{(q+r)^2} = - \frac{1}{(4 \pi)^2}
\frac{1}{6}
\end{eqnarray}
While $a_2$ depends on the choice of the cutoff function $K$, the
constant $C$ is independent.

\section{Change of field variables}

In modifying ERG differential equations, we use linear changes of
field variables as the main tool.  This has been discussed in detail
for the four dimensional theory in ref.~\cite{Sonoda:2003um}.

\subsection{First type}

We introduce the following infinitesimal change of field variables:
\begin{equation}
\phi (p) \to \phi (p) \left( 1 + \frac{1}{2} s (p) \right)
\end{equation}
where $s(p)$ is given by
\begin{equation}
s (p) \equiv - \delta z + (1 - K(p)) \left( \delta z + \frac{\delta
  m^2}{p^2 + m^2 \e^{2 t}} \right)
\end{equation}
Both $\delta z$ and $\delta m^2$ are infinitesimal constants.  Under
this change of variables, the squared mass changes as
\begin{equation}
m^2 \to m^2 + \delta m^2
\end{equation}
and the vertices change as
\begin{eqnarray}
\delta \V_2 (t; p, -p) &=& \delta z (p^2 + m^2 \e^{2t}) + \delta m^2
\nonumber\\
&& \quad + (1 + s(p))\, \V_2 (t; p, -p)\\
\delta \V_{2n \ge 4} (t; p_1, \cdots, p_{2n}) &=& \frac{1}{2}
\sum_{i=1}^{2n} s(p_i) \cdot \V_{2n} (t; p_1, \cdots, p_{2n})
\end{eqnarray}
Obviously, $\delta m^2$ and $\delta z$ are the counterterms for the
squared mass and wave function, respectively.  It is this type of
change of variables that we will use in
sects.~\ref{self-similarity}\&\ref{WF} to modify the ERG differential
equations.

\subsection{Second type}

With $u(p)$ as an arbitrary infinitesimal function of $p^2$, we
introduce the change of variables:
\begin{equation}
\phi (p) \to \phi (p) \left( 1 + \frac{1}{2} u (p) \right)
\end{equation}
This changes the propagator as
\begin{equation}
\frac{K(p)}{p^2 + m^2 \e^{2t}} \to \frac{K(p)}{p^2 + m^2 \e^{2t}}
\left( 1 - u (p) \right)
\end{equation}
and the vertices as
\begin{equation}
\V_{2n} (t; p_1,\cdots, p_{2n}) \to \left( 1 + \frac{1}{2}
\sum_{i=1}^{2n} u (p_i) \right) \V_{2n} (t; p_1, \cdots, p_{2n})
\end{equation}

If we assume that $u(p)$ is local, i.e.,
\begin{equation}
u(p) = 0 \quad \mathrm{if}\quad p^2 < 1
\end{equation}
then we can absorb the change of the propagator by changing the
vertices as follows:
\[
\delta \V_{2n} (t; p_1, \cdots, p_{2n}) = \frac{1}{2}
\sum_{i=1}^{2n} u(p_i) \cdot \V_{2n} (t; p_1, \cdots,
p_{2n})
\]
\parbox{4.5cm}{\hfill$\displaystyle - \sum_{\mathrm{partitions}}$}
\parbox{1.5cm}{\epsfig{file=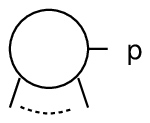}}
\parbox{2cm}{$\displaystyle \frac{K(p) u(p)}{p^2 + m^2 \e^{2t}}$}
\parbox{1.5cm}{\epsfig{file=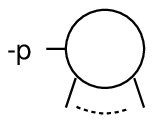}}
\begin{equation}
 - \frac{1}{2} \int_q \frac{K(q) u(q)}{q^2 + m^2 \e^{2t}}
\V_{2(n+1)} (t; q, -q, p_1, \cdots, p_{2n})
\end{equation}

The most general linear change of field variables is obtained by
combining the first and second types.

\subsection{Unphysical nature of the parameter $b_2$}

Now we are ready to discuss the unphysical nature of the parameter
$b_2$ in detail.  We combine the first and second types with the
choice
\begin{eqnarray}
\delta m^2 &=& \ep \lambda^2 \e^{2t}\\ \delta z &=& 0\\ u (p \e^t) &=&
u(t; p) \equiv \left( 1 - K(p \e^t) \right) \frac{\ep \lambda^2}{p^2 +
m^2}
\end{eqnarray}
so that
\begin{equation}
s (p \e^t) = u (t; p)
\end{equation}
Then, we obtain the following infinitesimal change of vertices:
\begin{eqnarray}
&&\e^{- 2t} \delta \V_2 (t; p \e^t, - p \e^t) = \ep \lambda^2 + 2 u
  (t; p) \cdot \e^{- 2 t} \V_2 (t; p \e^t, -p \e^t) \nonumber\\
&& \quad - \e^{- 2t} \V_2 (t; p \e^t, - p \e^t) \frac{ K(p \e^t) u (t;p)
  }{p^2 + m^2} \e^{- 2t} \V_2 (t; p \e^t, - p \e^t)\nonumber\\
&& \quad - \frac{1}{2} \int_q \frac{K (q \e^t) u (t;q)}{q^2+m^2} \e^{- t}
  \V_4 (t; q \e^t, - q \e^t, p \e^t, - p \e^t)
\end{eqnarray}
and
\[
 \e^{- y_{2n} t} \delta \V_{2n \ge 4} (t; p_1 \e^t, \cdots, p_{2n}
  \e^t)\\
= \sum_{i=1}^{2n} u (t; p_i) \cdot \e^{- y_{2n} t} \V_{2n} (t;
  p_1 \e^t, \cdots, p_{2n} \e^t)
\]
\parbox{3cm}{\hfill$\displaystyle - \sum_{\mathrm{partitions}}$}
\parbox{1.5cm}{\epsfig{file=vertexp.eps}}
\parbox{2.3cm}{$\displaystyle \frac{K (p \e^t) u (t;p)}{p^2 + m^2}$}
\parbox{1.5cm}{\epsfig{file=pvertex.eps}}
\begin{equation}
- \frac{1}{2} \int_q \frac{K (q \e^t) u(t;q)}{q^2 + m^2} \,
  \e^{- y_{2(n+1)} t} \V_{2(n+1)} (t; q \e^t, - q \e^t, p_1 \e^t,
  \cdots, p_{2n} \e^t)
\end{equation}

The above change of variables is very special in the sense that the
modified vertices $\{ (\V_{2n} + \delta \V_{2n})(t)\}$ satisfy the
same ERG differential equations as $\{\V_{2n} (t)\}$ except that the
squared mass parameter $m^2$ is replaced by
\begin{equation}
m^2 + \delta m^2 = m^2 + \ep \lambda^2
\end{equation}
It is straightforward (but tedious) to check this.

Since the vertices $\{ (\V_{2n} + \delta \V_{2n})(t)\}$ are obtained
from $\{\V_{2n} (t)\}$ by the change of field variables
\begin{equation}
\phi (p \e^t) \longrightarrow \phi (p \e^t) \left( 1 + u (t; p) \right)
\end{equation}
the correlation functions do not change \footnote{Strictly
  speaking, we must restrict $p_i^2  < 1$ for all $i$.}:
\begin{equation}
\vev{\phi (p_1) \cdots \phi (p_{2n})}_{m^2 \e^{2t}; \V (t)} =
\vev{\phi (p_1) \cdots \phi (p_{2n})}_{(m^2 + \ep \lambda^2) \e^{2t};
(\V + \delta \V) (t)}
\end{equation}
Examining the change of the parameters $\lambda$ and $b_2$ under the
above infinitesimal change
\begin{eqnarray}
\e^{- 2t} \delta \V_2 (t; p \e^t, - p \e^t) &\stackrel{t \to -
  \infty}{\longrightarrow}& \ep \lambda^2\\
\e^{-t} \delta \V_4 (t; p_1 \e^t, \cdots, p_4 \e^t) &\stackrel{t \to -
  \infty}{\longrightarrow}& 0
\end{eqnarray}
we obtain
\begin{equation}
\delta b_2 = \ep,\quad \delta \lambda = 0
\end{equation}
Hence, we find
\begin{eqnarray}
&&(\V_{2n} + \delta \V_{2n}) (t; p_1, \cdots, p_{2n}; m^2, \lambda,
  b_2)\nonumber\\
&&= \V_{2n} (t; p_1, \cdots, p_{2n}; m^2 + \ep \lambda^2, \lambda, b_2 +
\ep)
\end{eqnarray}
Therefore, the theory parametrized by $m^2, \lambda,~b_2$ gives the
same correlation functions as the theory with $m^2 + \ep \lambda^2,
\lambda,~b_2 + \ep$:
\begin{equation}
(m^2, \lambda, b_2)
  \stackrel{\mathrm{equivalent}}{\Longleftrightarrow} (m^2 + \ep
  \lambda^2, \lambda, b_2 + \ep) \label{equiv}
\end{equation}
This shows the unphysical nature of the parameter $b_2$.  For
instance, we can adopt the convention
\begin{equation}
b_2 = 0
\end{equation}
since, given an arbitrary ERG trajectory with $b_2 \ne 0$, we can
always find an equivalent ERG trajectory satisfying this condition.
\begin{figure}
\begin{center}
\epsfig{file=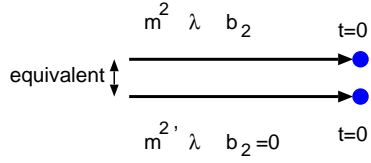}
\caption{Given a trajectory with $b_2 \ne 0$, we can find an
  equivalent trajectory satisfying $b_2 = 0$. (${m^2}' = m^2 - b_2
  \lambda^2$)}
\end{center}
\end{figure}

\subsection{RG equations from ERG}

Using the convention $b_2 = 0$, the ERG trajectories are now specified
only by $m^2$ and $\lambda$.  We wish to derive the RG equations for $m^2$
and $\lambda$ from ERG.

Let us recall the result (\ref{shift}).  Using this, we can shift the
logarithmic scale parameter $t$ by an infinitesimal $\Delta t$ in the
equivalence (\ref{equiv}):
\begin{equation}
(m^2 \e^{2 \Delta t}, \lambda \e^{\Delta t}, b_2 - C \Delta t)
\stackrel{\mathrm{equivalent}}{\Longleftrightarrow} 
(m^2 \e^{2 \Delta t} + \Delta t \cdot C \lambda^2, \lambda \e^{\Delta
  t}, b_2)
\end{equation}
where we have chosen $\ep = C \Delta t$.  This implies that the ERG
trajectory specified by $m^2, \lambda,~b_2$ gives the same correlation
functions as the trajectory specified by $m^2 \e^{2 \Delta t} + \Delta
t\cdot C \lambda^2, \lambda \e^{\Delta t},~b_2$, except that the
logarithmic scale parameter $t$ of the latter trajectory is shifted by
$\Delta t$.
\begin{figure}
\begin{center}
\epsfig{file=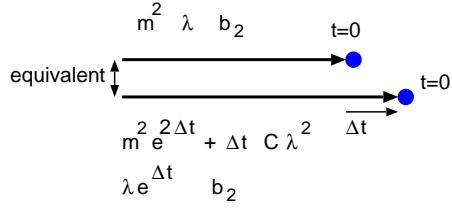}
\caption{The ERG trajectory specified by $m^2, \lambda,~b_2$ is
  equivalent with the trajectory specified by $m^2 \e^{2
\Delta t} + \Delta t\cdot C \lambda^2, \lambda \e^{\Delta t},~b_2$.
But the logarithmic scale parameter $t$ is shifted by $\Delta t$.}
\end{center}
\end{figure}
Thus, we obtain
\begin{eqnarray}
&&\vev{\phi (p_1 \e^{\Delta t}) \cdots \phi (p_{2n} \e^{\Delta
  t})}_{(m^2 \e^{2 \Delta t} + \Delta t \cdot C \lambda^2) \e^{2t}; \V
  (t; m^2 \e^{2 \Delta t} + \Delta t \cdot C \lambda^2, \lambda
  \e^{\Delta t}, b_2)}\nonumber\\ &=& \e^{(y_{2n} - 4 n) \Delta t}
  \vev{\phi (p_1) \cdots \phi (p_{2n})}_{m^2 \e^{2 t}; \V (t; m^2,
  \lambda, b_2)}
\end{eqnarray}
Taking $b_2 = 0$ and $t = 0$, we obtain
\begin{eqnarray}
&&\vev{\phi (p_1 \e^{\Delta t}) \cdots \phi (p_{2n} \e^{\Delta
  t})}_{m^2 \e^{2 \Delta t} + \Delta t \cdot C \lambda^2; \V
  (t=0; m^2 \e^{2 \Delta t} + \Delta t \cdot C \lambda^2, \lambda
  \e^{\Delta t}, 0)}\nonumber\\
&& = \e^{(y_{2n} - 4 n) \Delta t}
  \vev{\phi (p_1) \cdots \phi (p_{2n})}_{m^2 ; \V (t=0; m^2,
  \lambda, 0)}
\end{eqnarray}
This is the standard RG equation for the $\phi^4$ theory in three
dimensions.  The RG equations for the parameters $m^2, \lambda$ are
given by
\begin{eqnarray}
\frac{d m^2}{dt} &=& 2 m^2 + C \lambda^2\\
\frac{d \lambda}{dt} &=& \lambda
\end{eqnarray}
What is counterintuitive about the above result is that the ordinary
RG equations are obtained by comparing two \textbf{different} ERG
trajectories.

\section{Second rewriting: self-similarity \label{self-similarity}}

We have explained one undesirable feature of the ERG differential
equations: for self-similarity we must keep an unphysical parameter
$b_2$ in addition to the physical parameters $m^2, \lambda$.  The
purpose of this section is to modify the ERG differential equations so
that the solutions, parametrized only by two physical parameters, are
self-similar.

We modify the ERG differential equations so that the asymptotic
behavior of the two-point vertex is given by
\begin{equation}
\e^{- 2t} \V_2 (t; p \e^t, - p \e^t) \stackrel{t \to -
  \infty}{\longrightarrow} \lambda \e^{-t} a_2
\end{equation}
where $a_2$ is given by (\ref{a2}), without any terms of order $1$
proportional to $\lambda^2$.  We modify the ERG differential equations
by adding a mass counterterm as discussed in the previous section.
The modified ERG differential equations are given as follows:
\begin{eqnarray}
&& \frac{\partial}{\partial t} \left(\e^{- 2t} \V_2 (t; p \e^t, - p
  \e^t)\right) = - a_2 \lambda (t)^2 \e^{- 2t} \nonumber\\ && \quad+ s
  (p \e^t; m^2 (t), \lambda (t)) \cdot \e^{- 2t} \V_2 (t; p \e^t, - p
  \e^t) \nonumber\\ && \quad+ \e^{- 2t} \V_2 (t; p \e^t, - p \e^t)
  \frac{\Delta (p \e^t)}{p^2 + \e^{- 2t} m^2 (t)} \e^{- 2t} \V_2 (t; p
  \e^t, - p \e^t)\nonumber\\ && \quad + \frac{1}{2} \int_q
  \frac{\Delta (q \e^t)}{q^2 + \e^{-2t}m^2 (t)} \, \e^{- t} \V_4 (t; q
  \e^t, - q \e^t, p \e^t, - p \e^t), \\ &&\frac{\partial}{\partial t}
  \left( \e^{- y_{2n} t} \V_{2n \ge 4} (t; p_1 \e^t, \cdots, p_{2n}
  \e^t) \right) \nonumber\\ &=& \frac{1}{2} \sum_{i=1}^{2n} s (p_i
  \e^t; m^2 (t), \lambda (t)) \cdot \e^{- y_{2n} t} \V_{2n} (t; p_1
  \e^t, \cdots, p_{2n} \e^t)\nonumber
\end{eqnarray}
\parbox{3.5cm}{\hfill$\displaystyle + \sum_{\mathrm{partitions}}$}
\parbox{1.5cm}{\epsfig{file=vertexp.eps}}
\parbox{2.5cm}{$\displaystyle \frac{\Delta (p \e^t)}{p^2 +
  \e^{- 2t} m^2(t)}$}
\parbox{1.5cm}{\epsfig{file=pvertex.eps}}
\begin{equation}
+ \frac{1}{2} \int_q \frac{\Delta (q \e^t)}{q^2 + \e^{-2t}m^2 (t)} \, \e^{-
  y_{2(n+1)} t} \V_{2(n+1)} (t; q \e^t, -q \e^t, p_1 \e^t, \cdots,
  p_{2n} \e^t)
\end{equation}
where the running parameters are given by
\begin{eqnarray}
m^2 (t) &\equiv& \e^{2t} \left( m^2 + C \lambda^2 t \right)\\
\lambda (t) &\equiv& \e^t \lambda
\end{eqnarray}
and
\begin{equation}
s (p; m^2, \lambda) \equiv C \lambda^2 \frac{1 - K(p)}{p^2 + m^2}
\end{equation}

As in the case of the original ERG differential equations, the
solutions originating from the trivial fixed point at $t = - \infty$
are completely characterized by the asymptotic behaviors, which are
given in this case as follows:
\begin{eqnarray}
\e^{- 2t} \V_2 (t; p \e^t, - p \e^t) &\stackrel{t \to -
  \infty}{\longrightarrow}& \e^{-t} \lambda a_2\\
\e^{- t} \V_4 (t; p_1 \e^t, \cdots, p_4 \e^t)  &\stackrel{t \to -
  \infty}{\longrightarrow}& - \lambda\\
\e^{- y_{2n} t} \V_{2n \ge 4} (t; p_1 \e^t, \cdots, p_{2n} \e^t)  &\stackrel{t \to -
  \infty}{\longrightarrow}& 0
\end{eqnarray}
We can construct integral equations that incorporate the above
asymptotic behaviors.  For the two-point vertex we obtain
\begin{eqnarray}
&&\e^{- 2 t} \V_2 (t; p \e^t, - p \e^t)\nonumber\\
&=& \int_{-\infty}^t dt' \Bigg[ \e^{- 2t'} \V_2 (t'; p \e^{t'}, - p
    \e^{t'}) \frac{\Delta (p \e^{t'})}{p^2 + \e^{- 2 t'} m^2 (t')}
  \e^{- 2t'} \V_2 (t'; p \e^{t'}, - p
    \e^{t'}) \nonumber\\
&& + \frac{1}{2} \int_q \frac{\Delta (q \e^{t'})}{q^2 + \e^{- 2 t'}
      m^2 (t')} \e^{- t'} \V_4 (t'; q \e^{t'}, - q \e^{t'}, p \e^{t'},
    - p \e^{t'})\nonumber\\
&& + \e^{- t'} \lambda a_2 + C \lambda^2 + s (p \e^{t'}; m^2 (t'),
    \lambda (t')) \e^{- 2t'} \V_2 (t'; p \e^{t'}, - p \e^{t'}) \,
    \Bigg]\nonumber\\ 
&& + \e^{-t} \lambda a_2
\end{eqnarray}
For the four- and higher-point vertices, we obtain

\vspace{0.3cm}
\parbox{6cm}{\hspace{0.6cm}$\displaystyle
\e^{- y_{2n} t} \V_{2n \ge 4} (t; p_1 \e^t,
\cdots, p_{2n} \e^t)$}\\
\parbox{3.5cm}{$\displaystyle = \int_{-\infty}^t dt' \Bigg[\,
    \sum_{\mathrm{partitions}}$}
\parbox{3cm}{\epsfig{file=erg1.eps}}
\parbox{1cm}{$\displaystyle + \frac{1}{2} \int_q$}
\parbox{2cm}{\epsfig{file=erg2.eps}}\hfill
\begin{equation}
+ \frac{1}{2} \sum_{i=1}^{2n} s (p_i \e^{t'}; m^2 (t'), \lambda (t'))
 \cdot \e^{- y_{2n} t'} \V_{2n} (t'; p_1 \e^{t'}, \cdots, p_{2n}
 \e^{t'}) \,\Bigg] \,- \lambda \delta_{n,2}
\end{equation}
where the thick line with momentum $q$ denotes
\[
\frac{\Delta (q \e^{t'})}{q^2 + \e^{- 2t'} m^2 (t')}
\]
By shifting the integration variable $t'$ by $t$ so that the range of
integration becomes $[-\infty, 0]$, we find that the vertices are
indeed self-similar:
\begin{equation}
\fbox{$\displaystyle \V_{2n} (t; p_1, \cdots, p_{2n}) = F_{2n} (p_1,
\cdots, p_{2n}; m^2 (t), \lambda (t))$}
\end{equation}
where $F_{2n}$ has no explicit $t$ dependence.

Thus, we have accomplished the goal of this section, and the ERG flows
now coincide with the RG flows of the running parameters $m^2 (t)$ and
$\lambda (t)$.  The RG equations are the same as those derived at the
end of the previous section, and they are identical to the RG
equations for the dimensionally regularized theory with the minimal
subtraction.  

However, there is a problem with the above RG equations: the
Wilson-Fisher fixed point is hidden at infinite $\lambda$.  Let us
define an RG invariant
\begin{equation}
R (m^2, \lambda) \equiv \frac{m^2}{\lambda^2} - C \ln \lambda
\end{equation}
which takes a critical value, say $R_{cr}$, for the massless theory.
Then, the Wilson-Fisher fixed point lies at the infinite $\lambda$
limit of the critical RG trajectory $R = R_{cr}$.  In order to obtain
the fixed point at finite values of parameters, we need yet another
modification of the ERG differential equations.  This is the subject
of the next section.

\section{Third rewriting: Wilson-Fisher fixed point \label{WF}}

Assuming self-similarity, we write our $2n$-point vertex as
\[
\V_{2n} (p_1, \cdots, p_{2n}; m^2, \lambda)
\]
By introducing counterterms for the squared mass and wave function, we
wish to modify the differential equations so that the following
conditions are met:
\begin{eqnarray}
\V_2 (0,0; 0, \lambda) &=& a_2 \lambda \label{v2}\\
\frac{\partial}{\partial m^2} \V_2 (0,0; m^2, \lambda)\Big|_{m^2=0}
&=& 0 \label{v2m}\\
\frac{\partial}{\partial p^2} \V_2 (p, -p; 0, \lambda)\Big|_{p^2=0}
&=& 0 \label{v2p}
\end{eqnarray}
Note that these are not asymptotic conditions.  The first and second
conditions determines the mass counterterm, and the third the wave
function renormalization.  We define $\lambda$ by
\begin{equation}
\V_4 (0,0,0,0; 0, \lambda) = - \lambda \label{v4}
\end{equation}

The modified ERG differential equations are given as follows:
\begin{eqnarray}
&&\frac{\partial}{\partial t} \left( \e^{- 2 t} \V_2 (p \e^t, - p
\e^t; m^2 (t), \lambda (t)) \right) \nonumber\\ &=& \e^{- 2t}
\left\lbrace \beta_m (\lambda (t)) m^2 (t) + \eta (\lambda (t)) \left(
p^2 \e^{2t} + m^2 (t) \right) \right\rbrace\nonumber\\ && \, + s (p
\e^t; m^2 (t), \lambda (t)) \cdot \e^{- 2t} \V_2 (p \e^t, - p \e^t;
m^2 (t), \lambda(t))\nonumber\\ &&\, + \left\lbrace \e^{- 2t} \V_2 (p
\e^t, - p \e^t; m^2 (t), \lambda (t)) \right\rbrace^2 \frac{\Delta (p
\e^t)}{p^2 + \e^{- 2t} m^2 (t)} \nonumber\\ && + \frac{1}{2} \int_q
\frac{\Delta (q \e^t)}{q^2 + \e^{- 2t} m^2 (t)} \e^{- t} \V_4 (q \e^t,
- q \e^t, p \e^t, - p \e^t; m^2 (t), \lambda (t)), \\ &&
\frac{\partial}{\partial t} \left( \e^{- y_{2n} t} \V_{2n} (p_1 \e^t,
\cdots, p_{2n} \e^t; m^2 (t), \lambda (t) )\right) \nonumber\\ &=&
\frac{1}{2} \sum_{i=1}^{2n} s( p_i \e^t; m^2 (t), \lambda (t)) \cdot
\e^{- y_{2n} t} \V_{2n} (p_1 \e^t, \cdots, p_{2n} \e^t; m^2 (t),
\lambda (t) \nonumber
\end{eqnarray}
\parbox{3cm}{\hfill$\displaystyle + \sum_{\mathrm{partitions}}$}
\parbox{2.5cm}{\epsfig{file=erg1.eps}}
\parbox{1cm}{$\displaystyle + \frac{1}{2} \int_q$}
\parbox{2.5cm}{\epsfig{file=erg2.eps}}\\
where the thick line with momentum $q$ denotes
\[
\frac{\Delta (q \e^t)}{q^2 + \e^{- 2t} m^2 (t)}
\]
In the above, $\beta_m (\lambda)$ is the anomalous dimension of $m^2$
so that
\begin{equation}
\frac{d}{dt} m^2 (t) = ( 2 + \beta_m (\lambda (t)) ) m^2 (t) + c
(\lambda (t))
\end{equation}
and $\frac{1}{2} \eta (\lambda)$ is the anomalous dimension of the
scalar field.  The function $s$ is defined by
\begin{equation}
s (p; m^2, \lambda) \equiv - \eta (\lambda) K (p) + \left( \beta_m
(\lambda) m^2 + c (\lambda) \right) \frac{1 - K(p)}{p^2 + m^2}
\end{equation}

The beta function $\beta (\lambda)$, defined as usual by
\begin{equation}
\frac{d}{dt} \lambda (t) = \beta (\lambda (t)),
\end{equation} 
and the other three functions $\beta_m (\lambda), c (\lambda), \eta
(\lambda)$ are determined so that the above ERG differential equations
satisfy the four conditions (\ref{v2}-\ref{v4}).  Using the notation
\begin{eqnarray}
A_{2n} (p_1, \cdots, p_{2n}; \lambda) &\equiv& \V_{2n} (p_1, \cdots,
p_{2n}; 0, \lambda)\\
B_{2n} (p_1, \cdots, p_{2n}; \lambda) &\equiv&
\frac{\partial}{\partial m^2} \V_{2n} (p_1, \cdots, p_{2n}; m^2, \lambda)
\Big|_{m^2=0}\\
C_{2n} (p_1, \cdots, p_{2n}; \lambda) &\equiv&
\frac{\partial^2}{(\partial m^2)^2}  \V_{2n} (p_1, \cdots, p_{2n};
m^2, \lambda)\Big|_{m^2=0}
\end{eqnarray}
we find
\begin{eqnarray}
&& a_2 ( \beta (\lambda) - \lambda + \lambda \eta (\lambda)) -
c(\lambda) = \frac{1}{2} \int_q \frac{\Delta (q)}{q^2} \left( A_4
(q,-q,0,0;\lambda) + \lambda \right)\label{c}\\ &&\beta_m (\lambda) + \eta
(\lambda) - c (\lambda ) C_2 (0,0;\lambda) \nonumber\\ &&\qquad\qquad =
\frac{1}{2} \int_q \Delta (q) \left( - \frac{B_4
(q,-q,0,0;\lambda)}{q^2} + \frac{A_4 (q,-q,0,0;\lambda)}{q^4}
\right)\label{betam}\\ &&\eta (\lambda) - c(\lambda) \frac{\partial}{\partial p^2}
B_2 (p, -p; \lambda)\Big|_{p^2=0} \nonumber\\ && \qquad\qquad = - \frac{1}{2}
\frac{\partial}{\partial p^2} \int_q \frac{\Delta (q)}{q^2} A_4
(q,-q,p,-p;\lambda) \Big|_{p^2=0}\label{eta}\\&& \beta (\lambda) - \lambda + 2
\lambda \eta (\lambda) - c(\lambda) B_4 (0,0,0,0;\lambda) \nonumber\\
&&\qquad\qquad = - \frac{1}{2} \int_q \frac{\Delta (q)}{q^2} A_6
(q,-q,0,0,0,0;\lambda)\label{beta}
\end{eqnarray}
These imply
\begin{eqnarray}
\beta (\lambda) - \lambda &=& \mathrm{O} (\lambda^2)\\
\beta_m (\lambda) &=& \mathrm{O} (\lambda)\\
c(\lambda) &=& \mathrm{O} (\lambda^2)\\
\eta (\lambda) &=& \mathrm{O} (\lambda^2)
\end{eqnarray}
and we can introduce the following series expansions:
\begin{eqnarray}
\beta (\lambda) - \lambda &=& \beta_1 \lambda^2 + \beta_2 \lambda^3 + \cdots\\
\beta_m (\lambda) &=& \beta_{m1} \lambda + \beta_{m2} \lambda^2 +
\cdots\\
c(\lambda) &=& c_2 \lambda^2 + \cdots\\
\eta (\lambda) &=& \eta_2 \lambda^2 + \cdots
\end{eqnarray}

The Wilson-Fisher point $({m^2}^*, \lambda^*)$ is found from
\begin{eqnarray}
\beta (\lambda^*) &=& 0\\
(2 + \beta_m (\lambda^*)) {m^2}^* + c (\lambda^*) &=& 0
\end{eqnarray}
where $\beta_m^* \equiv \beta_m (\lambda^*)$ and $\eta^* \equiv \eta
(\lambda^*)$ are the anomalous dimensions of the squared mass and
scalar field, respectively.  Both should be independent of the choice
of the cutoff function $K$, but we have not been able to demonstrate
the independence.  If we truncate the series expansions in $\lambda$,
both $\beta_m^*$ and $\eta^*$ depend on the choice of $K$ as we will
see later.

\subsection{Integral equations}

For perturbative calculations in powers of $\lambda$, we have found it
convenient to convert the ERG differential equations and the four
conditions (\ref{v2} - \ref{v4}) into integral equations.  For the
two-point vertex, we obtain
\begin{eqnarray}
&& \e^{- 2t} \V_2 (p \e^t, -p \e^t; m^2 (t), \lambda (t))\nonumber\\
&=& \int_{-\infty}^t dt' \Bigg[ \beta_m (\lambda (t')) \e^{- 2t'} m^2
    (t') + \eta (\lambda (t')) \left( p^2 + \e^{- 2t'} m^2
    (t')\right)\nonumber\\
&&\, + s (p \e^{t'}; m^2 (t'), \lambda (t')) \e^{- 2t'} \V_2 (p
    \e^{t'}, - p \e^{t'}; m^2 (t'), \lambda (t'))\nonumber\\
&&\qquad + \eta (\lambda (t')) a_2 \e^{- 2t'} \lambda (t') \nonumber\\
&& \, + \left( \e^{- 2t'} \V_2 (p \e^{t'}, - p \e^{t'}; m^2 (t'),
 \lambda (t')) \right)^2 \frac{\Delta ( p \e^{t'})}{p^2 + \e^{- 2 t'}
   m^2 (t')}\nonumber\\
&& \, + \frac{1}{2} \int_q \Delta (q \e^{t'}) \Bigg(
\frac{\e^{-t'} \V_4 (q \e^{t'}, - q \e^{t'}, p \e^{t'}, - p \e^{t'};
  m^2 (t'), \lambda (t'))}{q^2 + \e^{- 2t'} m^2 (t')}\nonumber\\
&&\qquad  - \frac{\e^{-
    t'} A_4 (q \e^{t'}, - q \e^{t'}, 0, 0; \lambda (t'))}{q^2}
\Bigg) \Bigg] \, + \e^{- 2 t} a_2 \lambda (t) 
\end{eqnarray}
For the four-point vertex, we obtain
\begin{eqnarray}
&& \e^{- t} \V_4 (p_1 \e^t, \cdots, p_4 \e^t; m^2 (t), \lambda
  (t))\nonumber\\
&=& \int_{-\infty}^t dt' \Bigg[ \frac{1}{2} \sum_{i=1}^4 s (p_i
  \e^{t'}; m^2 (t'), \lambda (t')) \cdot \e^{- t'} \V_4 (p_1 \e^{t'},
  \cdots, p_4 \e^{t'}; m^2 (t'), \lambda (t'))\nonumber\\
&&\, - 2 \eta (\lambda (t')) \e^{- t'} \lambda (t') + c (\lambda (t'))
  \e^{- t'} B_4 (0,0,0,0; \lambda (t'))\nonumber\\
&& \, + \sum_{i=1}^4 \e^{- 2t'} \V_2 (p_i \e^{t'}, - p_i \e^{t'}; m^2
  (t'), \lambda (t')) \frac{\Delta (p_i \e^{t'})}{p_i^2 + \e^{- 2 t'}
  m^2 (t')} \nonumber\\
&&\qquad\qquad \times \e^{- t'} \V_4 (p_1 \e^{t'}, \cdots, p_4 \e^{t'};
  m^2 (t'), \lambda (t')) \nonumber\\
&& \, + \frac{1}{2} \int_q \Delta (q \e^{t'}) \Big\lbrace \frac{\V_6
  (q \e^{t'}, - q \e^{t'}, p_1 \e^{t'}, \cdots, p_4 \e^{t'}; m^2 (t'),
  \lambda (t'))}{q^2 + \e^{- 2 t'} m^2 (t')}\nonumber\\
&&\qquad - \frac{A_6 (q \e^{t'}, - q \e^{t'}, 0,0,0,0; \lambda
  (t'))}{q^2} \Big\rbrace \Bigg] \, - \e^{-t} \lambda (t)
\end{eqnarray}
For the six- and higher-point vertices, we obtain
\begin{eqnarray*}
&& \e^{- y_{2n} t} \V_{2n \ge 6} (p_1 \e^t, \cdots, p_{2n} \e^t; m^2
  (t), \lambda (t)) = \int_{-\infty}^t dt' \Bigg[ \nonumber\\
&& \, \frac{1}{2} \sum_{i=1}^{2n} s (p_i
  \e^{t'}; m^2 (t'), \lambda (t')) \cdot \e^{- y_{2n} t'} \V_{2n} (p_1
  \e^{t'}, \cdots, p_{2n} \e^{t'}; m^2 (t'), \lambda (t'))
\end{eqnarray*}
\parbox{3cm}{\hfill$\displaystyle + \sum_{\mathrm{partitions}}$}
\parbox{2.5cm}{\epsfig{file=erg1.eps}}
\parbox{1.5cm}{$\displaystyle + \frac{1}{2} \int_q$}
\parbox{2cm}{\epsfig{file=erg2.eps}}
\parbox{1cm}{$\displaystyle \Bigg]$}\\
where the thick line with momentum $q$ denotes
\[
\frac{\Delta (q \e^{t'})}{q^2 + \e^{- 2t'} m^2 (t')}
\]
It is straightforward to check that the above integral equations give
the correct $t$-dependence and satisfy the conditions (\ref{v2} -
\ref{v4}).

\subsection{Results of perturbative calculations}

We will not give any details of the perturbative calculations, and
write down only the relevant results.  (See Appendix A for some
details.)

Using the series expansions of $\beta (\lambda)$, we find the fixed
point
\begin{equation}
\lambda^* \simeq \frac{1}{- \beta_1} \label{lambdastar}
\end{equation}
At lowest non-trivial order, the anomalous dimensions are obtained as
\begin{eqnarray}
\beta_m^* &\simeq& \frac{\beta_{m1}}{- \beta_1}\\
\eta^* &\simeq& \frac{\eta_2}{\beta_1^2}
\end{eqnarray}
All the coefficients are given in terms of the cutoff function $K$.
For example, if we choose
\begin{equation}
K(q) \equiv \left\lbrace \begin{array}{c@{\quad\mathrm{for}\quad}l}
 1 & q^2 < 1\\
 \frac{a^2 - q^2}{a^2-1} & 1 < q^2 < a^2\\
 0 & q^2 > a^2\end{array}\right.
\end{equation}
\begin{figure}
\begin{center}
\epsfig{file=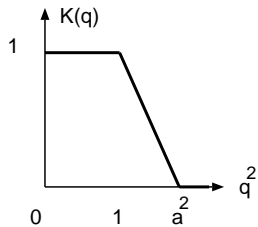}
\caption{A choice for the cutoff function.}
\end{center}
\end{figure}
(Fig.~4) then we obtain
\begin{eqnarray}
\beta_1 &=& - 3 \int_q \frac{\Delta (q)(1 - K(q))}{q^4} = -
\frac{1}{\pi^2} \frac{a+2}{(a+1)^2}\\
\beta_{m1} &=& - \frac{1}{2} \int_q \frac{\Delta (q)}{q^4} = -
\frac{1}{\pi^2} \frac{1}{2(a+1)}
\end{eqnarray}
so that the one-loop result
\begin{equation}
\beta_m^* \simeq \frac{\beta_{m1}}{- \beta_1} =  - \frac{1+a}{2(2+a)}
\end{equation}
is obtained.  This is between $- \frac{1}{3}~(a=1)$ and $-
\frac{1}{2}~(a \to \infty)$, comparable to the best fit $- 0.41$ to
various experimental results.  (For example, see Table 5.4.2 of
\cite{CMT}.)  With the above choice for $K$, we plot the $a$
dependence of the critical exponents $\beta_m^*$ at one-loop and
$\eta^*$ at two-loop (Fig.~5).
\begin{figure}
\begin{center}
\epsfig{file=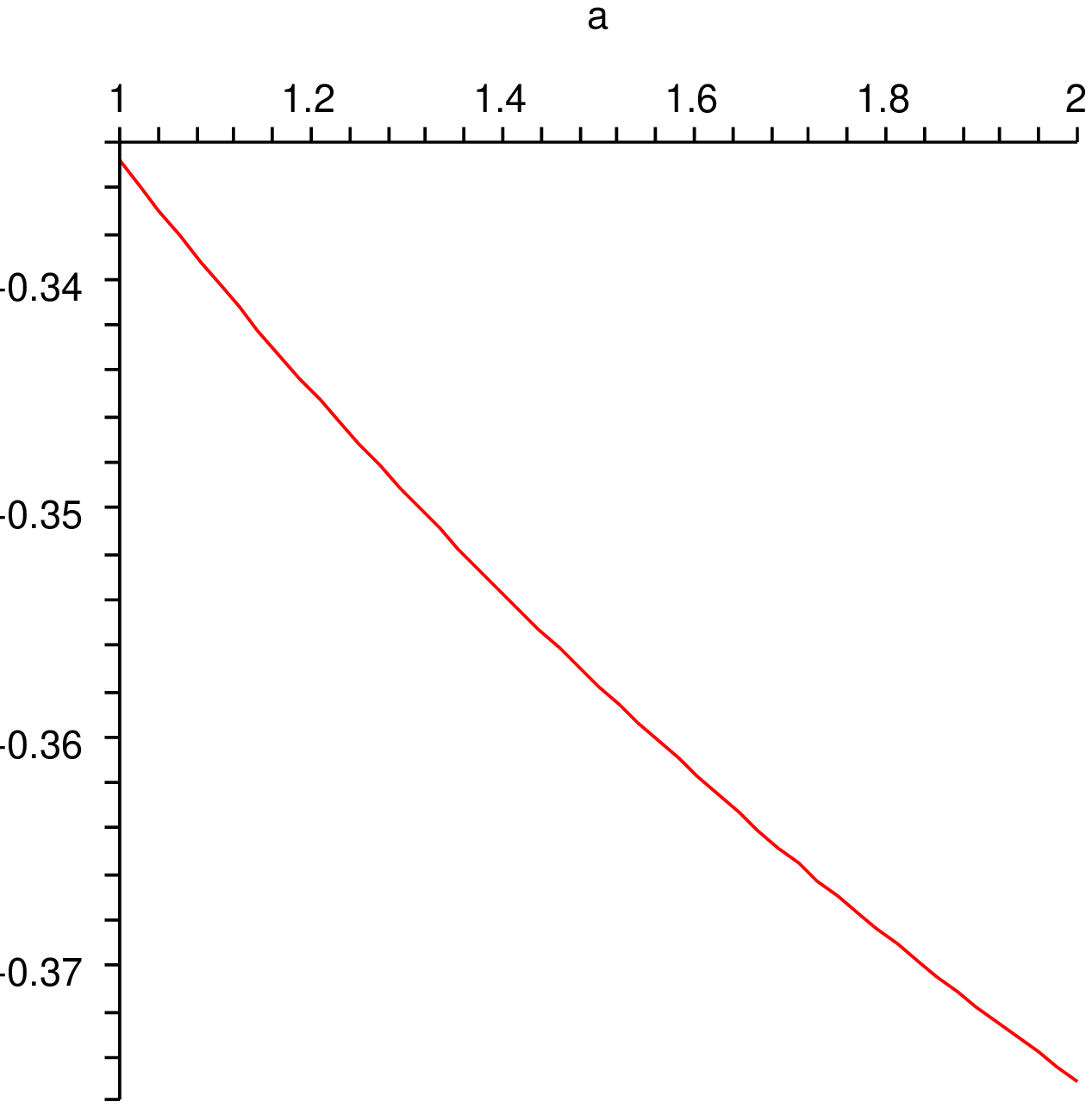, height=6cm} \epsfig{file=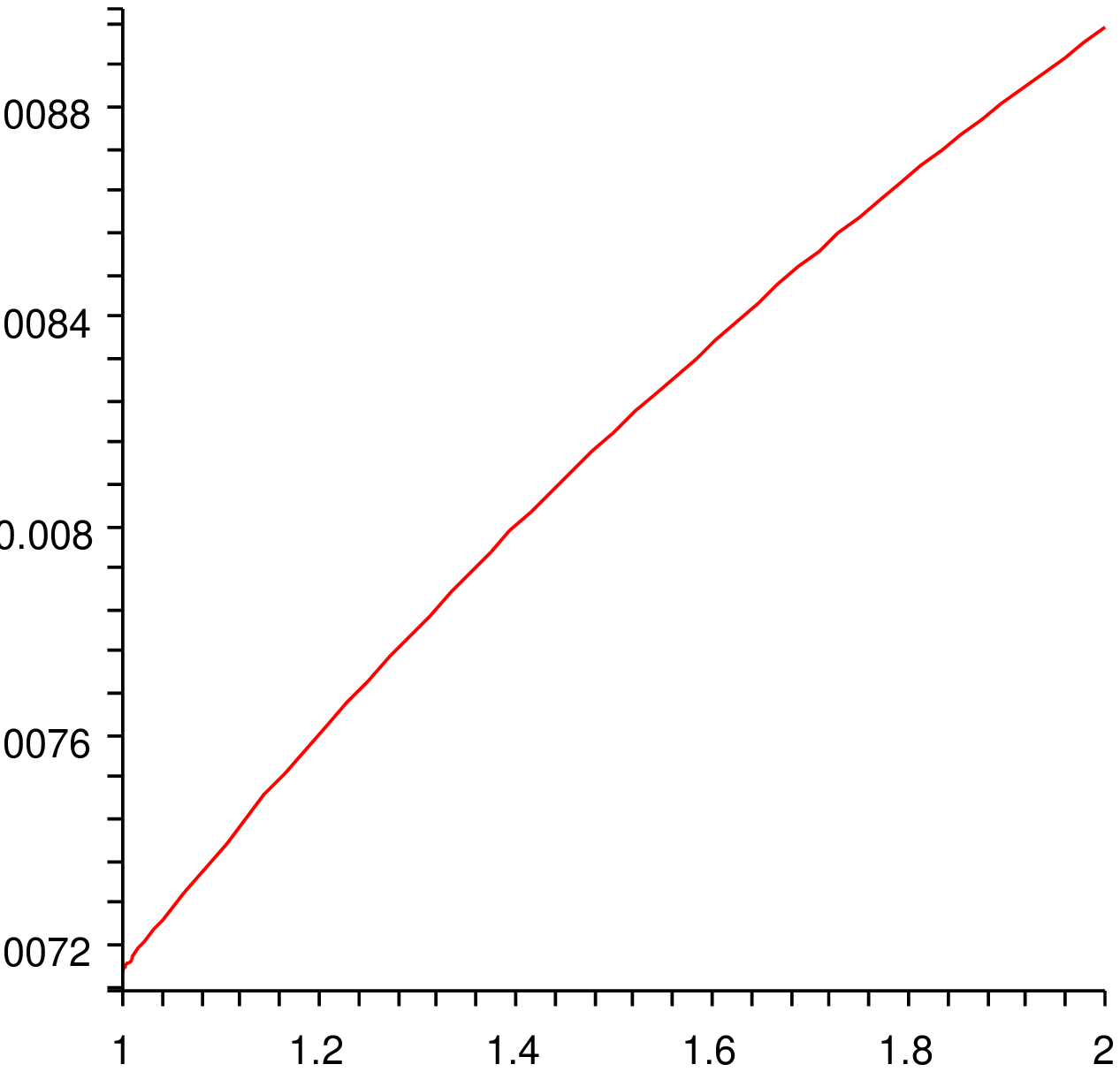, height=6cm}
\caption{$\beta_m^*$ and $\eta^*$ for $1 < a < 2$.}
\end{center}
\end{figure}
Our value for $\eta^*$ turns out to be too small compared to the
experimental fit $0.03-0.06$.  

In the limit $a \to 1+$, our formalism is expected to be equivalent to
the Wegner-Houghton ERG equations.\cite{Wegner:1972ih}  But we have
not yet examined this expected equivalence.

\section{Conclusion}

In this paper we have carefully examined the nature of the solutions
to Polchinski's ERG differential equations.  We have shown how the
ordinary RG equations of renormalized parameters arise from ERG.  We
have also shown the necessity to modify the ERG differential
equations, first for self-similarity, and second for the Wilson-Fisher
fixed point.  In all this analysis, we have found it helpful to use
the integral equation approach.

Our perturbative calculations of the critical exponents are
reminiscent of Parisi's use of the Callan-Symanzik equations to do the
same.\cite{Parisi} One undesirable feature common in both is the
absence of an obvious expansion parameter for the critical exponents.
Notice that $\lambda^*$ given by Eq.~(\ref{lambdastar}) is not
necessarily a small number.  Introducing $N \gg 1$ number of fields is
an easy way to rectify the problem, but it evades the question of
validity of perturbation theory presented in this paper.

The perturbative method given in this paper is by no means the only
way to calculate the critical exponents of the Wilson-Fisher fixed
point.  Typically the calculations are done non-perturbatively after
truncating the ERG equations.  See, for example,
ref.~\cite{Blaizot:2005xy} and references therein.

\appendix

\section{Results of perturbative expansions}

Using the notation
\begin{equation}
A_{2n} (p_1,\cdots,p_{2n}; \lambda) = \sum_{k=1}^\infty \lambda^k A_4^{(k)}
(p_1,\cdots,p_{2n})
\end{equation}
we obtain the following results:
\begin{eqnarray}
A_4^{(1)} (q,-q,0,0) &=& -1\\
A_6^{(2)} (q,-q,0,0) &=& 6 \frac{1 - K(q)}{q^2}
\end{eqnarray} 
at one-loop, and
\begin{eqnarray}
&&A_4^{(2)} (q,-q,0,0) = - 2 a_2 \frac{1 - K(q)}{q^2} \nonumber\\
&&\quad + \int_r \left( \frac{1 - K(r)}{r^2} \frac{1 - K(q+r)}{(q+r)^2} -
\frac{(1-K(r))^2}{r^4} \right)\\
&&B_4^{(2)} (q,-q,0,0) = \beta_{m1} \frac{\int_{-\infty}^0 dt \e^{t}
  (1 - K(q \e^t))}{q^2} + 2 a_2 \frac{1 - K(q)}{q^4}\nonumber\\
&&\quad - \int_r \frac{(1
  - K(r))^2}{r^6} - 2 \int_r \frac{1 - K(r)}{r^4} \frac{1 -
  K(q+r)}{(q+r)^2}\\
&&\frac{\partial}{\partial p^2} A_4^{(2)} (q,-q,p,-p)\Big|_{p^2=0} 
= \frac{\partial}{\partial p^2} \int_r \frac{1 - K(r)}{r^2} \frac{1 -
  K(p+q+r)}{(p+q+r)^2}\Big|_{p^2=0}\\
&&A_6^{(3)} (q,-q,0,0,0,0)\nonumber\\
&&\quad= 18 a_2 \frac{(1 - K(q))^2}{q^4} - 8
\beta_1 \frac{1 - K(q)}{q^2} - 3 \int_r \frac{\Delta (r) (1 -
  K(r))^2}{r^6}\nonumber\\
&&\qquad - 12 \frac{1 - K(q)}{q^2} \int_r \frac{1 - K(r)}{r^2} \frac{1
  - K(q+r)}{(q+r)^2} \nonumber\\
&&\qquad - 12 \int_r \frac{(1-K(r))^2 (1 - K(q+r))}{r^4 (q+r)^2}
\end{eqnarray}
at two-loop.

Now, from Eqs.~(\ref{c}-\ref{beta}) we obtain
\begin{eqnarray}
\beta_{m1} &=& \frac{1}{2} \int_q \frac{\Delta (q)}{q^4} A_4^{(1)}
(q,-q,0,0)
= - \frac{1}{2} \int_q \frac{\Delta (q)}{q^4}\\
\beta_1 &=& - \frac{1}{2} \int_q \frac{\Delta (q)}{q^2} A_6^{(2)}
(q,-q,0,0,0,0) = - 3 \int_q \frac{\Delta (q) (1 - K(q))}{q^4}
\end{eqnarray}
at one-loop, and
\begin{eqnarray}
a_2 \beta_1 - c_2 &=& \frac{1}{2} \int_q \frac{\Delta (q)}{q^2} A_4^{(2)}
(q,-q,0,0)\\
\beta_{m2} + \eta_2 &=& \frac{1}{2} \int_q \Delta (q) \left( -
\frac{B_4^{(2)} (q,-q,0,0)}{q^2} + \frac{A_4^{(2)}(q,-q,0,0)}{q^4}
\right)\label{betam2}\\
\eta_2 &=& - \frac{1}{2} \frac{\partial}{\partial p^2} \int_q
\frac{\Delta (q)}{q^2} A_4^{(2)} (q,-q,p,-p)\Big|_{p^2=0}\\
\beta_2 + 2 \eta_2 &=& - \frac{1}{2} \int_q \frac{\Delta (q)}{q^2}
A_6^{(3)} (q,-q,0,0,0,0)\label{beta2}
\end{eqnarray}
at two-loop, where
\begin{equation}
a_2 = \frac{1}{2} \int_q \frac{\Delta (q)}{q^2} 
\end{equation}
Hence, we get
\begin{eqnarray}
c_2 &=& - \frac{1}{2} \int_{q,r} \frac{1 - K(q)}{q^2} \frac{1 -
  K(r)}{r^2} \frac{\Delta (q+r)}{(q+r)^2} = - \frac{1}{6} \frac{1}{(4
  \pi)^2}\\
\eta_2 &=& - \frac{1}{2} \frac{\partial}{\partial p^2} \int_{q,r}
  \frac{\Delta (q)}{q^2} \frac{1 - K(r)}{r^2} \frac{1 -
  K(p+q+r)}{(p+q+r)^2} \Big|_{p^2=0}
\end{eqnarray}
Note that $c_2$ is the same as $C$ given at the end of
sect.~\ref{integral} and independent of the choice of $K$.  We can
also obtain $\beta_2, \beta_{m2}$ from (\ref{betam2}, \ref{beta2}) by
using the calculated four-point vertices.  In the main text we have
quoted the results for the critical exponents using a particular
cutoff function $K$ with one parameter $1 < a < 2$.

\bibliography{RG2005}

\end{document}